# Multifractal detrended fluctuation analysis of temperature in Spain (1960-2019)

Javier Gómez-Gómez [*], Rafael Carmona-Cabezas, Ana B. Ariza-Villaverde, Eduardo Gutiérrez de Ravé, Francisco José Jiménez-Hornero

GEPENA Research Group, University of Cordoba, Gregor Mendel Building (3rd floor), Campus Rabanales, 14071 Cordoba, Spain

[*] Corresponding author. E-mail address: f12gogoj@uco.es

**ABSTRACT**

In the last decades, an ever-growing number of studies are focusing on the extreme weather conditions related to the climate change. Some of them are based on multifractal approaches, such as the Multifractal Detrended Fluctuation Analysis (MF-DFA), which has been used in this work. Daily diurnal temperature range (DTR), maximum, minimum and mean temperature from five coastal and five mainland stations in Spain have been analyzed. For comparison, two periods of 30 years have been considered: 1960-1989 and 1990-2019. By using the MF-DFA method, generalized Hurst exponents and multifractal spectra have been obtained. Outcomes corroborate that all these temperature variables have multifractal nature and show changes in multifractal properties between both periods. Also, Hurst exponents values indicate that all time series exhibit long-range correlations and a stationary behavior. Coastal locations exhibit in general wider spectra for minimum and mean temperature than for maximum and DTR, in both periods. On the contrary, the mainland ones do not show this pattern. Also, width from multifractal spectra of these two variables is shortened in the last period in almost every case. To authors' mind, changes in multifractal features might be related to the climate change experienced in the studied region. Furthermore, reduction of spectra width for minimum and mean temperature implies a decrease of the complexity of these temperature variables between both studied periods. Finally, the wider spectra found in coastal stations might be useful as a discriminator element to improve climate models.

## 1. INTRODUCTION

For decades, it has been widely known the fact that air temperatures are increasing (on different spatial and time scales), as it has been proven by a number of different studies [1–3]. All of the last three decades have been characterized by being consecutively the warmest since 1850 [4], in terms of Earth's surface temperature. It is obvious that this temperature rise has a negative impact on live on Earth: changes in migration and number of many species; increase of the susceptibility of numerous ecosystems and human environments; great impacts to crops (wheat or maize among others). All these complications have an added negative influence on global politics, society and demographics [5].

Due to all this, there is an increasing interest in the scientific community regarding climate variability. The main approaches consist in climate models and statistical analyses that investigate extreme episodes which are supposed to be related to global warming. In these cases, the highest confidence levels are usually associated to unusually extreme cold and heat events [6]. Thus, the study of temperature variables time series is a widespread approach in the field [7–11].

In the case of Spain (which is where this work is focused), all findings seem to point to the fact that the main change in temperature during the twentieth century was recorded from the 1970s onward. This change was characterized by an abrupt and remarkable increase in temperatures. Nonetheless, this warming does not have a marked continuity nor regularity throughout the century. Neither it has been along the year, being winters typically when the strongest changes were identified. Also, several studies have shown that this change has been more pronounced for the maximum than for the minimum temperature in the Iberian Peninsula and in some subregions [12–15]. However, the opposite was found in other researches [16,17].

Classical statistical methods have been widely used traditionally in order to gain information from time series and to confirm climate models [18,19]. Furthermore, in the last decade, several advanced techniques have gain importance in the context of analysis of complexity and non-linearity of signal. Some of them are the so-called fractal and multifractal analysis [20–22]. Among these last ones, Multifractal Detrended Fluctuation Analysis (MF-DFA) [23] has become an extensively used technique for analyzing climatic time series [7,24–27]. This technique combines the Detrended Fluctuation Analysis (DFA) with the fractal theory, providing a reliable tool that yields information about complex and non-linear time series. DFA is used to determine fractal properties of non-stationary time series. However, it fails when it comes to characterizing series with more than one scaling exponent (multifractal), which is why MF-DFA has advantages over the first one.

The objective pursued in the presented work is to seek evidence of the influence of climate change in the multifractal properties of temperature time series in Spain. Furthermore, these multifractal properties will be analyzed to gain information on the nature and dynamics of the temperature time series. For such purpose, authors have selected four variables related to temperature (daily maximum, minimum and average temperature and the diurnal temperature range) at ten different locations across Spain and for two different time periods: 1960-1989 and 1990-2019. The employed technique for this analysis is the MF-DFA.

## 2. MATERIALS AND METHODS
### 2.1. Data

The studied data in this document correspond to four temperature time series of two periods of 30 years: 1960-1989 and 1990-2019. The data that support the findings of this study are openly available. They are provided by the Spanish Meteorological Agency ("Agencia Estatal de Meteorología") from the AEMET OpenData website at

[http://www.aemet.es/es/datos_abiertos/AEMET_OpenData](http://www.aemet.es/es/datos_abiertos/AEMET_OpenData). The four temperature variables are daily maximum ($T_{max}$), minimum ($T_{min}$) and mean temperature ($T_{max}$), and the diurnal temperature range ($DTR$), which is computed from the maximum and minimum temperatures (see Fig. 1). Raw data are recorded at 10 different meteorological stations located over the Iberian Peninsula, in Spain (Fig. 2). Half of stations belongs to coastal regions and the rest are mainland. Furthermore, they cover the Atlantic and the Mediterranean semiarid climates. Descriptive statistic shows a global increase of temperature variables, especially for mean temperature.

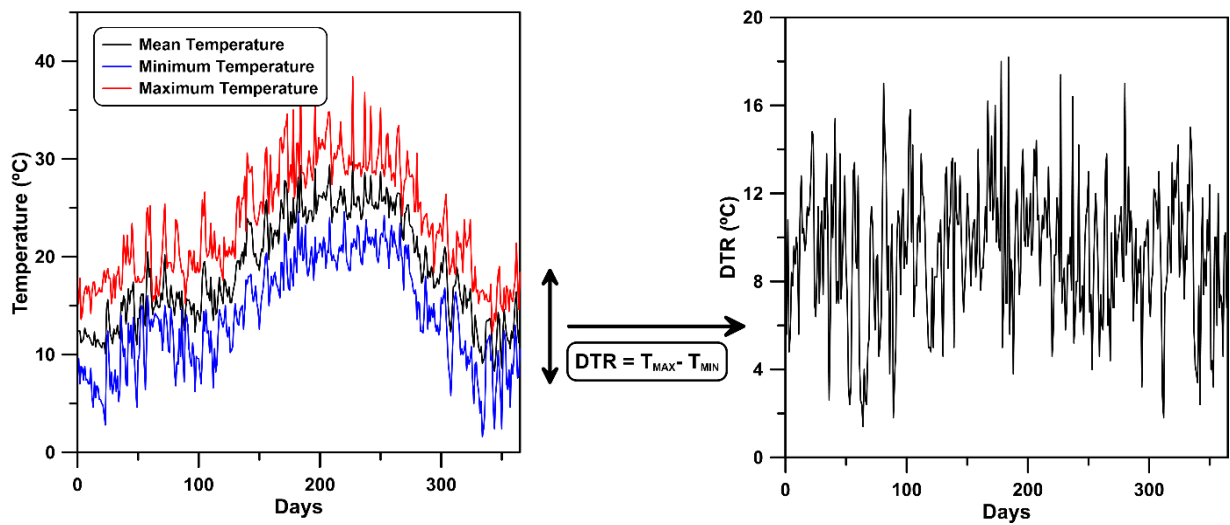

Fig. 1: Example of temperature variables records for 1990 in Málaga station. Daily mean, minimum and maximum temperature are shown on the left part, while diurnal temperature range ($DTR$) is depicted on the right. This last variable was directly computed by subtracting minimum temperature values to the maximum temperature.

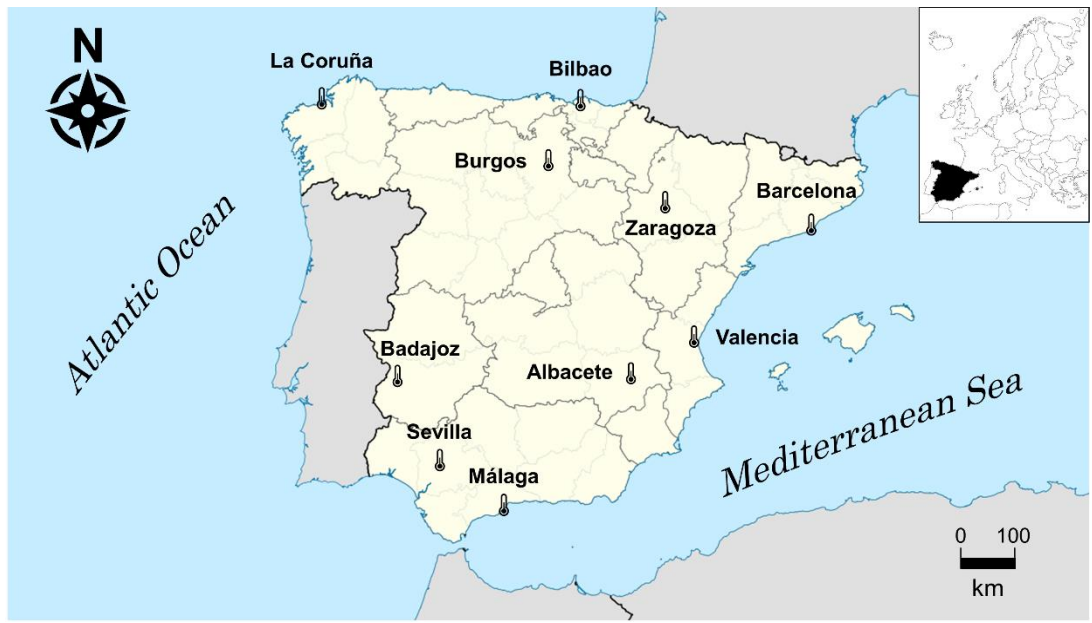

Fig. 2: Map of the studied meteorological stations localized in the Iberian Peninsula, Spain.

Before applying the MF-DFA, data must be preprocessed in order to remove the seasonality from time series. To this aim, average month values of temperature time series over all 30 years for each period have been computed. Next, these mean values are subtracted from the original signals to obtain the deseasonalized ones. To check that correlations due to seasonal effects are eliminated, authors have computed the autocorrelation functions of the original and the deseasonalized time series. In Fig. 3, an example of these functions is depicted. As it can be observed, the autocorrelation functions before this procedure is applied, give an almost sinusoidal behavior in the interval $[-1, +1]$ which soften when increasing scale. After month values are subtracted, these functions decay rapidly to zero.

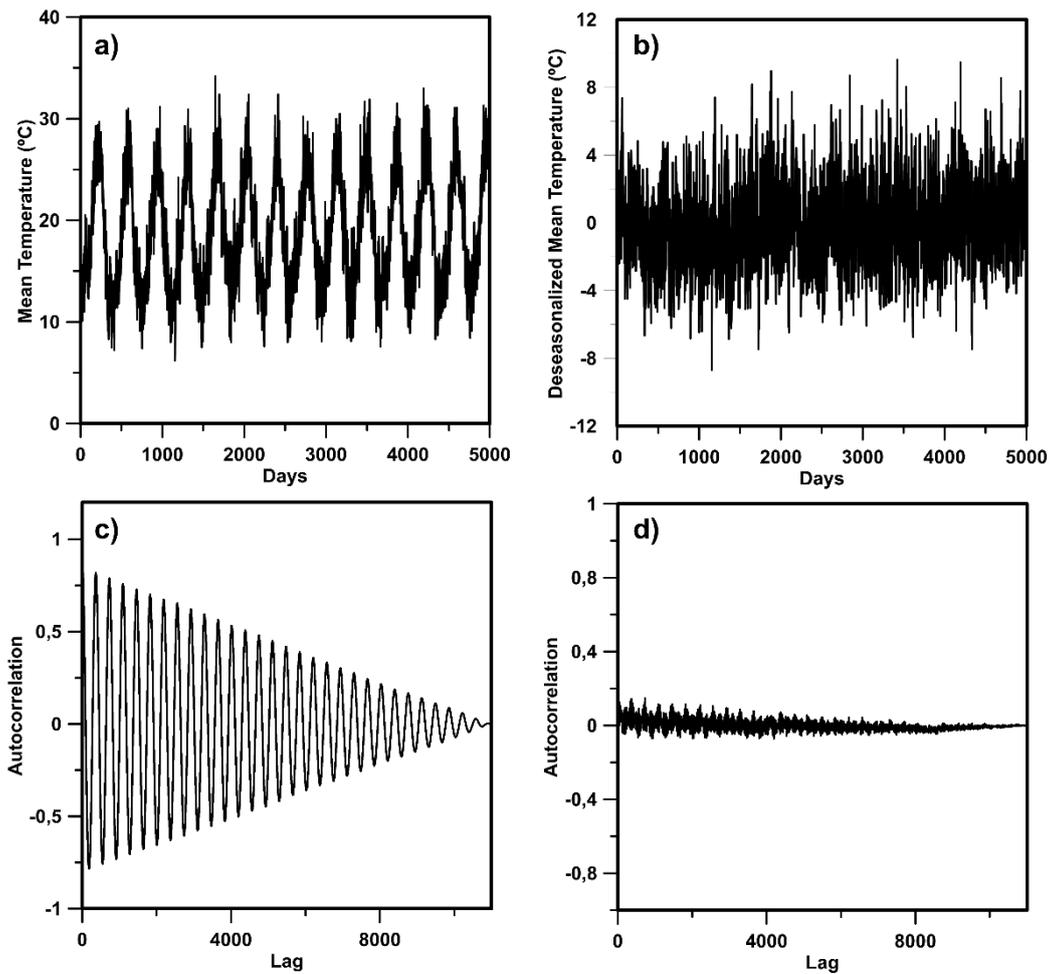

Fig. 3: Example of original (a) and deseasonalized (b) mean temperature time series for Málaga station in the period 1990-2019 and their corresponding autocorrelations functions (c and d, respectively). For clarity reasons, time series are plotted only for the first 5000 data (or days).

## 2.1. Multifractal Detrended Fluctuation Analysis

MF-DFA was a method proposed by Kantelhardt et al. [28] for multifractal analysis of nonstationary time series and it is based on detrended fluctuation analysis (DFA). The main advantage of these both approaches is that they can obtain the scaling behavior of the fluctuations in time series. Although the studied system is affected by artificial correlations derived from unknown underlining trends, these techniques allow to retrieve the intrinsic fluctuations of the system [29]. DFA was invented in order to deal with monofractal time series and the main concepts were extended by Kantelhardt et al. to multifractal signals. The five steps to implement the MF-DFA algorithm are the following [28]:

1) Firstly, compute the integrated time series, also known as the "profile". Let $x_k$ be a time series of length $N$ and $\langle x \rangle$ the mean value. Then, the profile is defined as:

$$Y(i) \equiv \sum_{k=1}^{i} [x_k - \langle x \rangle], \qquad i = 1, \dots, N \tag{1}$$

2) Next, divide the profile $Y(i)$ into $N_s \equiv \text{int}(N/s)$ nonoverlapping segments of equal length $s$. The length $N$ of the series is often not a multiple of the time scale $s$, thus, a short part at the end of the profile may remain. To hold this part, the same procedure is repeated from the end of the series to the beginning. Thereby, $2N_s$ segments are obtained for each time scale $s$.

3) Compute the local trend for each segment $v$ by means of the least-squares fit of the series. The fitting polynomial $y_v(i)$ can be linear, quadratic, cubic, or higher order polynomial. Different order of the polynomial fit differs in the capability to eliminate trends in the series [30]. In $m$th order of MF-DFA, trends or order $m$ in the profile (or, equivalently, of order $m - 1$ in the original series) are eliminated. Therefore, by subtracting $y_v(i)$ for each segment, one can compute the variance for each $s$ value:

$$F^2(v, s) \equiv \frac{1}{s} \sum_{i=1}^{s} \{Y[(v - 1)s + i] - y_v(i)\}^2 \tag{2}$$

for each segment $v$, $v = 1, \ldots, N_s$ and

$$F^2(v,s) \equiv \frac{1}{s}\sum_{i=1}^{s}\{Y[N-(v-N_s)s+i]-y_v(i)\}^2 \qquad (3)$$

for each segment $v$, $v = N_s + 1, \ldots, 2N_s$.

4) Average over all segments to obtain the $q$th order fluctuation function:

$$F_q(s) \equiv \left\{\frac{1}{2N_s}\sum_{v=1}^{2N_s}[F^2(v,s)]^{q/2}\right\}^{1/q} \qquad (4)$$

where the index $q$ can take any real value except zero, because of the diverging exponent. For $q = 0$, a logarithmic averaging procedure must be performed and the fluctuation function is computed as follows:

$$F_{q=0}(s) \equiv \exp\left\{\frac{1}{4N_s}\sum_{v=1}^{2N_s}\ln[F^2(v,s)]\right\} \qquad (5)$$

For $q = 2$, the standard DFA method is obtained. To retrieve the scaling behavior of the generalized $q$ dependent fluctuation functions, steps 2 to 4 must be repeated for different time scales $s$. $F_q(s)$ will increase with increasing $s$. Besides, this fluctuation function depends on the $m$ order of the polynomial fit and, by construction, is only defined for $s \geq m + 2$ [30].

5) To determine the scaling behavior of the fluctuation functions, it is necessary to analyze the log-log plots of $F_q(s)$ vs $s$ for each value of $q$. If the series $x_k$ is long-range power-law correlated, then $F_q(s)$ increases for larges values of $s$ as a power law:

$$F_q(s) \sim s^{h(q)} \qquad (6)$$

Hence, the scaling exponent $h(q)$ can be computed by obtaining the slopes of the log-log plots of $F_q(s)$ vs $s$ for each $q$. For very large scales, $F_q(s)$ becomes statistically inaccurate for the averaging procedure, since the number of segments $2N_s$ becomes very small. Also, systematic deviations from the scaling behavior occur for very small scales ($s \approx 10$). Thus, a thorough analysis is needed in order to determine the best range for the least-squares fits.

In general, $h(q)$ can depend on $q$. For stationary time series, $h(2)$ is the well-known Hurst exponent $H$ whereas, for non-stationary signals, the Hurst exponent is $H = h(2) - 1$ [31]. For this reason, $h(q)$ is called as generalized Hurst exponent. On the other hand, the Hurst exponent value ($H$) gives information about the correlation properties of the signals. For a white noise process (uncorrelated time series), $H = 0.5$. When $0 < H < 0.5$, the signal is anti- long-range anticorrelated, meaning that a large value is more likely to be followed by a small value and vice versa. Finally, when $H > 0.5$, time series is long-range correlated and large values are more likely to be followed by other large values and vice versa [29].

For monofractal time series, $h(q)$ is independent of $q$ and Eq. (4) gives an identical scaling behavior for all values of $q$. Only if small and large fluctuations scales differently, $h(q)$ will depend significantly on $q$. Segments with large variance $F^2(v, s)$ or large deviations from the fit will dominate the average value $F_q(s)$ for $q > 0$. On the contrary, segments with small variance $F^2(v, s)$ will dominate $F_q(s)$ for $q < 0$. Therefore, $h(q)$ describes the scaling behavior of the segments with large fluctuations (when $q > 0$) and the scaling behavior of the segments with small fluctuations (when $q < 0$).

## 2.2. Relation to Standard Multifractal Analysis

In order to relate MF-DFA method to the standard multifractal analysis based on the box counting formalism, Kantelhardt et al. also demonstrated that the scaling exponent $h(q)$ is related to the scaling exponent $\tau(q)$, which is defined by the partition function of the multifractal formalism [28]. This relationship is stablished by the expression:

$$\tau(q) = qh(q) - 1 \qquad (7)$$

Another way to characterize a multifractal series in the standard formalism is by means of the so-called multifractal spectrum or singularity spectrum $f(\alpha)$, which can be computed from $\tau(q)$ via the Legendre transform:

$$\alpha = \frac{d\tau(q)}{dq} \text{ and } f(\alpha) = q\alpha - \tau(q) \qquad (8)$$

where $\alpha$ is the singularity strength or Hölder exponent and the shape of $f(\alpha)$ is usually a concave-down parabola with a maximum value which correspond to the most dominant scaling behavior [25]. The corresponding value of the singularity strength at this maximum is often denoted by $\alpha_0$ and the width of the multifractal spectrum (i.e., $W = \alpha_{max} - \alpha_{min}$) gives information about the degree of the multifractality of the signal [32]. When the time series is monofractal, the width of the spectrum will be close to zero.

If the curve is fitted by a second order polynomial, it can be obtained an asymmetry parameter $B$ to discern between right-skewed or left-skewed distributions. Hence, the multifractal spectrum can be parametrized by:

$$f(\alpha) = A(\alpha - \alpha_0)^2 + B(\alpha - \alpha_0) + C \qquad (9)$$

When $B = 0$, the spectrum is symmetrical, whereas for $B > 0$ is left-skewed and for negative values is right-skewed [33,34]. A right-skewed spectrum is related to relatively strongly weighted high fractal exponents. Its broadness is mainly due to small fluctuations ($q < 0$) and the time series is more regular (with "fine-structure") [35,36]. On the contrary, a left-skewed spectrum indicates a relatively strongly weighted low fractal exponents associated to large fluctuations ($q > 0$) and a more singular signal. Thus, it shows a richer multifractal structure in the arrangement of the large fluctuations.

## 3. RESULTS AND DISCUSSION

### 3.1. Generalized Hurst Exponents

To calculate the generalized Hurst exponent $h(q)$ of temperature variables, fluctuation functions $F_q(s)$ with a range of $q$ values from -5 to 5 with step 0.5 has been chosen. The interval selected for the scale values $s$ is from 5 to 2000 days with step of 5 days. In Fig. 4, values of $\log[F_q(s)]$ vs $\log(s)$ is shown for Málaga in the period 1990-2019 as an example. It is observed that fluctuation functions increase with scale. As it can also be appreciated, curves can be fitted by a linear regression to obtain the sought generalized Hurst exponent $h(q)$. The optimal range to compute the linear fit is approximately between 18 and 355 days (almost one year). Nonetheless, in some cases it is necessary to shorten this range (down to 126 days) to avoid artifacts that worsen the analysis.

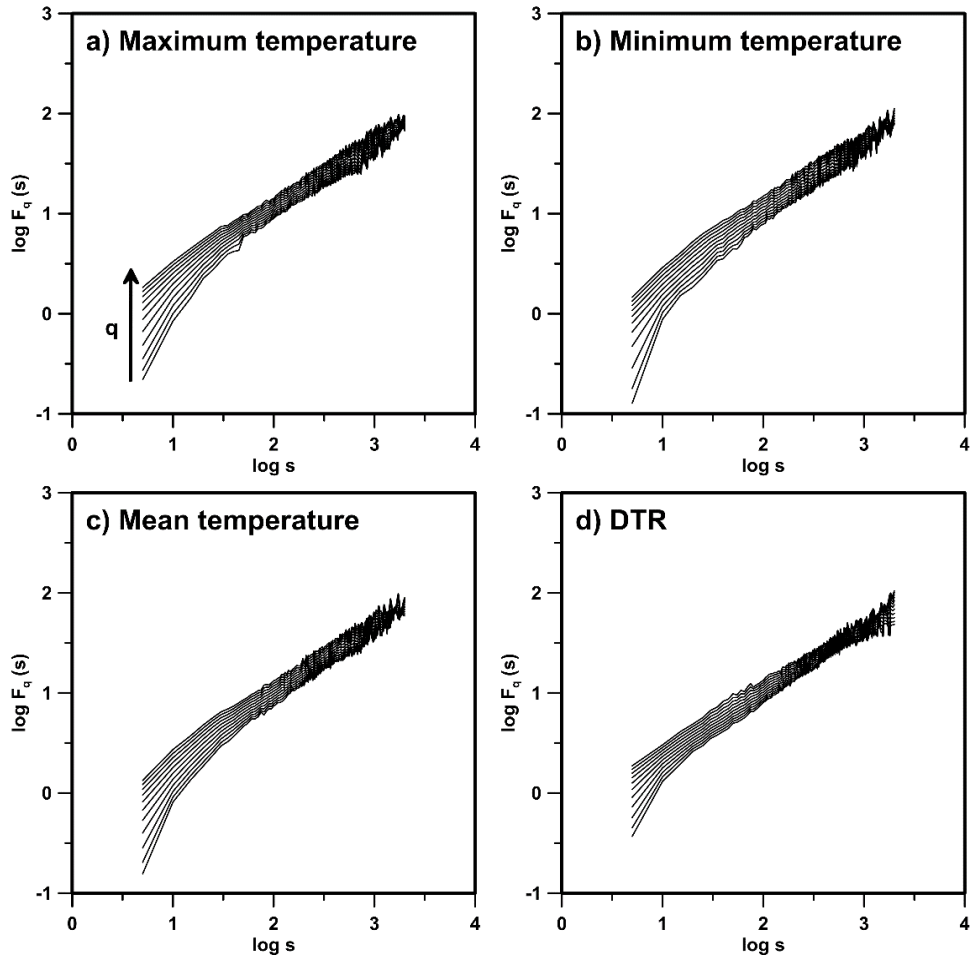

Fig. 4: Fluctuation functions vs scale values of segments for Málaga station in the period 1990-2019. Each curve corresponds to one $q$ value. To make clearer charts, only half of the analyzed curves are depicted.

The generalized Hurst exponent for every location and period of time has been retrieved from their respective least squares regressions and results are plotted in Fig. 5 for the five coastal locations and in Fig. 6 for the mainland ones.

On the one hand, looking at the five coastal stations, it can be observed that, in general, the scaling exponent is a decreasing function of $q$. One can define a quantity $\Delta h(q) = h_{max} - h_{min}$, being $h_{max}$ the maximum value of the generalized Hurst exponent in the interval and $h_{min}$ the minimum. The value of $\Delta h(q)$ gives information about the multifractality degree of the signal. Meaning that a greater multifractality degree is related to more violent temperature fluctuations [37]. $\Delta h(q)$ is higher for mean and minimum temperature variables. This fact indicates a more multifractal behavior from these signals which derives from more complex systems [23].

Furthermore, the only case where $\Delta h(q)$ is almost zero occurs for $DTR$ in Málaga station in the years 1960-1989 (Fig. 5g). This is a characteristic behavior of monofractal time series. Apparently, for the next 30 years, the $DTR$ evolves slightly to a more multifractal signal in this case, as shown in Fig. 5h. However, this will be clearer when multifractal spectra are discussed further in the text.

On the other hand, the mainland stations in Fig. 6 present a similar behavior such as the decreasing trend. Here, negligible differences in $\Delta h(q)$ are shown, contrary to the coastal locations where minimum and mean temperature had a more pronounced value of $\Delta h(q)$. Physically speaking, the major degree of multifractality for mean and minimum temperature in the coastal stations might be related to the oceanic influence. Looking at Zaragoza, a monofractal nature is identified as well in this case for DTR in the 1960-1989 period. This tendency changes for the next period, exhibiting a higher multifractal degree (same phenomenon happened with Málaga, as discussed before).

Overall, the Hurst exponent $H$ value can be calculated from these curves for $q = 2$. Because all these times series have a value of $h(2) < 1$, they are demonstrated to be stationary signals and the Hurst exponent is exactly this value [38]. All values of this parameter are in the range $[0.601, 0.777]$. As $H > 0.5$ for all the cases, time series are long-range correlated, meaning that a relative high value of signals are likely to be followed by other high value and vice versa [25]. Regarding the $\Delta h(q)$, it must be point out the fact that for the studied series, their values belong to the interval $[0.027, 0.187]$. This shows that there is a high variability of the multifractal degree among the different series: from almost monofractal (as seen with Málaga and Zaragoza) to clearly multifractal ones.

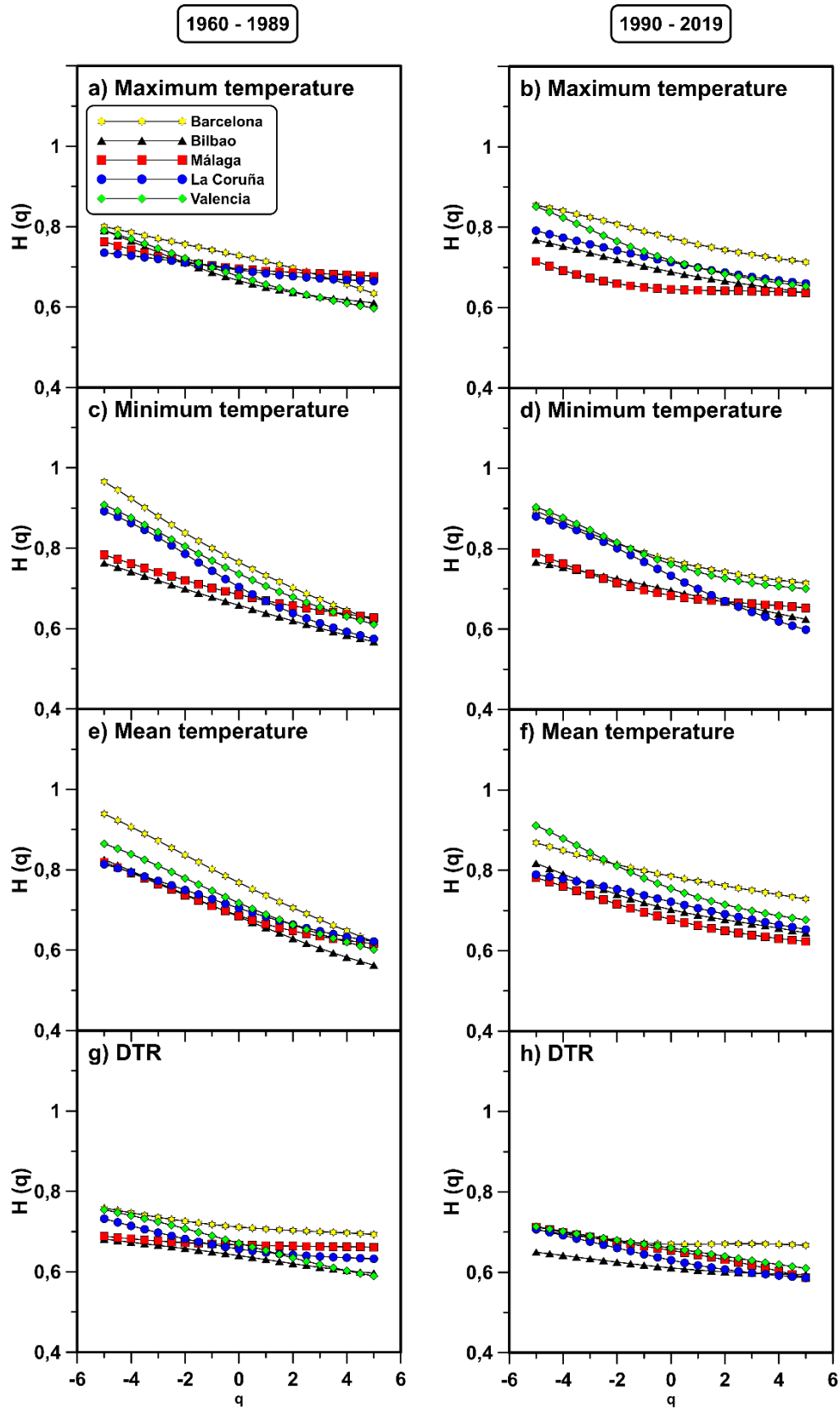

Fig. 5: Generalized Hurst exponents for $T_{max}$ (a and b), $T_{min}$ (c and d), $T_{mean}$ (e and f) and $DTR$ (g and h) in the five coastal stations (Barcelona, Bilbao, Málaga, La Coruña and Valencia). Charts on the left side are from the years 1960-1989 (a, c, e and g), while the ones on the right side correspond to the years 1990-2019.

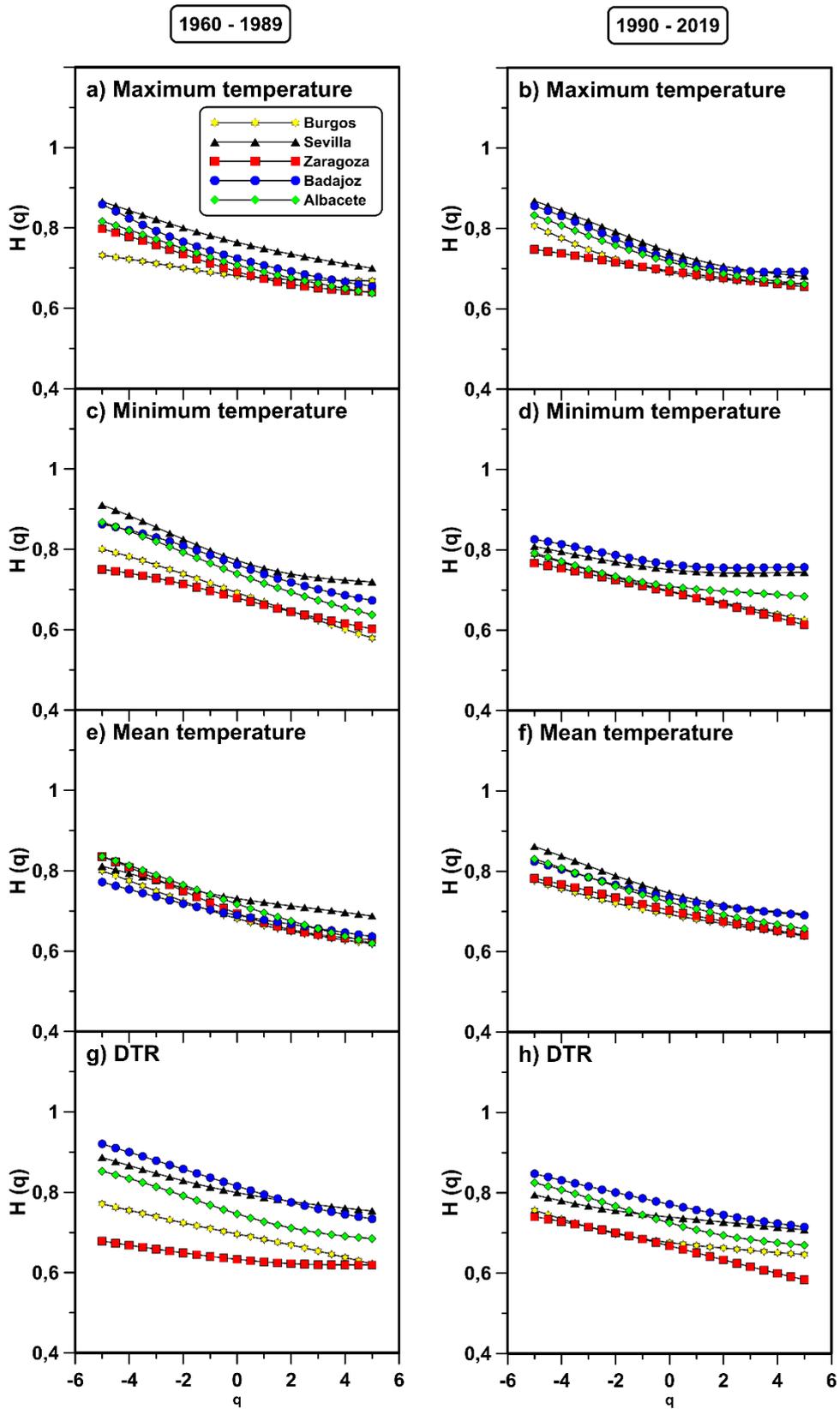

Fig. 6: Generalized Hurst exponents for $T_{max}$ (a and b), $T_{min}$ (c and d), $T_{mean}$ (e and f) and $DTR$ (g and h) in the five mainland stations (Burgos, Sevilla, Zaragoza, Badajoz and Albacete). Charts on the left side are from the years 1960-1989 (a, c, e and g), while the ones on the right side correspond to the years 1990-2019.

### 3.2. Multifractal Spectra

Multifractal spectra are obtained by means of the computed scaling exponent $\tau(q)$, which is yielded from the relation between this quantity and the generalized Hurst exponent (see Eq. (7)). Hölder exponents $\alpha$ and $f(\alpha)$ are finally retrieved from the Legendre transform of this scaling exponent. In Fig. 7 and Fig. 8, it can be seen the multifractal spectra for every city, period and temperature variable used in this analysis in the same order as Fig. 5 and Fig. 6. Next, the coastal stations will be discussed.

Looking at the maximum temperature variable in Figs. 7a and b, it can be observed that some differences are present between both periods. For all stations except for Málaga, the position of the maxima $\alpha_0$, which denotes the dominant singularity strength, is slightly shifted to the right from years 1960-1989 to 1990-2019. On the contrary, Málaga spectrum is shifted to the left, meaning that Málaga changes to more correlated signal and more regular structure in the last 30 years. Meanwhile, the other four do the opposite, becoming more complex signals [24]. When it comes to the width ($W = \alpha_{max} - \alpha_{min}$), La Coruña is the coastal station that changes the most between both periods, increasing the degree of multifractality (see Table 1). Barcelona and Valencia spectra have a rather shorter left tail for the last period in contrast to the first one, denoting that, in the last 30 years, there is more homogeneity in the large fluctuations for this series.

Minimum temperature signals (Figs. 7c and d) show that in this case the peaks of the spectra ($\alpha_0$) experience a shift to the right in the second time period (the series become less correlated, as explained before). The width of the spectra is in this case reduced for every case, pointing to a reduction of the multifractality of the minimum temperature time series over the years. Again, it is possible to see that the left tail corresponding to Barcelona and Valencia is shorter for 1990-2019, as happened with the maximum temperature.

For mean temperature (Figs. 7e and f), again $\alpha_0$ slightly moves to the right and the widths of the spectra are shortened for every location from years 1960-1989 to 1990-2019. Furthermore, the most highlighted stations that present a reduced left tail between both periods are Barcelona and Valencia again, being coherent with the previous results.

DTR results (Figs. 7g and h) also depict slight shifts in the value of $\alpha_0$. The difference in this case is that these changes are now towards the left direction of the x-axis. Now the changes in the value of $W$ are not as consistent as in the previous variables. Multifractal degree of La Coruña and Málaga are increased, while the others decrease. The last mentioned station (Málaga) stands out by having a much larger left tail for the last years.

In general, multifractal spectra for minimum and mean temperature are wider than in the other variables, meaning that the degree of multifractality of these variables is larger and that these time series have more complex behavior. This fact is coherent with the outcomes obtained from the generalized Hurst exponent. On the other hand, the asymmetry parameter $B$, which is shown in Table 2, indicates change in the sign of symmetry for several stations and variables. La Coruña and Valencia do not change the sign of $B$. Barcelona alters its symmetry for $T_{max}$ from negative to positive, which denotes that spectrum changes from right to left-skewed and becomes more singular in the last period, as discussed in the Sec. 2.2. Bilbao modify its symmetry from positive to negative in $T_{min}$, i.e., it becomes smoother or less singular for the last years. On the contrary, it is altered from negative to positive in spectra for $DTR$ (see Table 2). Lastly, Málaga changes from positive to negative in $T_{max}$ and $DTR$.

Once the discussion of the results for the coastal stations has been done, the equivalent for the mainland ones is described, which can be seen in Fig. 8. Focusing on the maximum temperature (Figs. 8a and b), it seems that there is no common behavior when it comes to the shift of $\alpha_0$ for the five stations. Regarding the shape and the width of the spectra, it must be pointed out that most of them are very similar, especially for the second period, except for Zaragoza.

Moving to the minimum temperature spectra (Figs. 8c and d), the $\alpha_0$ positions shifts vary from city to another. While for Sevilla and Albacete become more correlated (move to the left), the rest do the opposite. The width of the spectra decreases this time for all of them, except for Zaragoza, that remains almost the same. It can be clearly seen in the corresponding figure.

For mean temperature, in every mainland station it can be observed how spectra are shifted to the right (see Figs. 8e and f), meaning that signals become more complex. By looking at the width, it decreases in all locations except for Sevilla. Again, the shape of spectra is very alike, more notably for the last 30 years.

Lastly, $DTR$ charts (Figs. 8g and h) depict shifted spectra to the left in every case except for Zaragoza. For the width, almost all the locations show a decrease of the degree of multifractality. Again, Zaragoza has a different behavior, increasing the width instead of decreasing. Indeed, the spectrum changes from almost monofractal in the years 1960-1989 to multifractal in the last period. This agrees with the results of the generalized Hurst exponent.

Overall, it cannot be said that the minimum and mean temperature spectra are wider, as happened to the coastal stations. Hence, the multifractal degree in this case is relatively similar for all the variables. In this case, the asymmetry parameter B (see Table 2) maintains its sign for every city and variables, except for Sevilla ($T_{max}$) and Zaragoza ($T_{min}$ and $DTR$). In these cases, the sign always changes from positive to negative, which means as explained before, that the spectra change from left to right-skewed. Therefore, the signals become more regular in the last period.

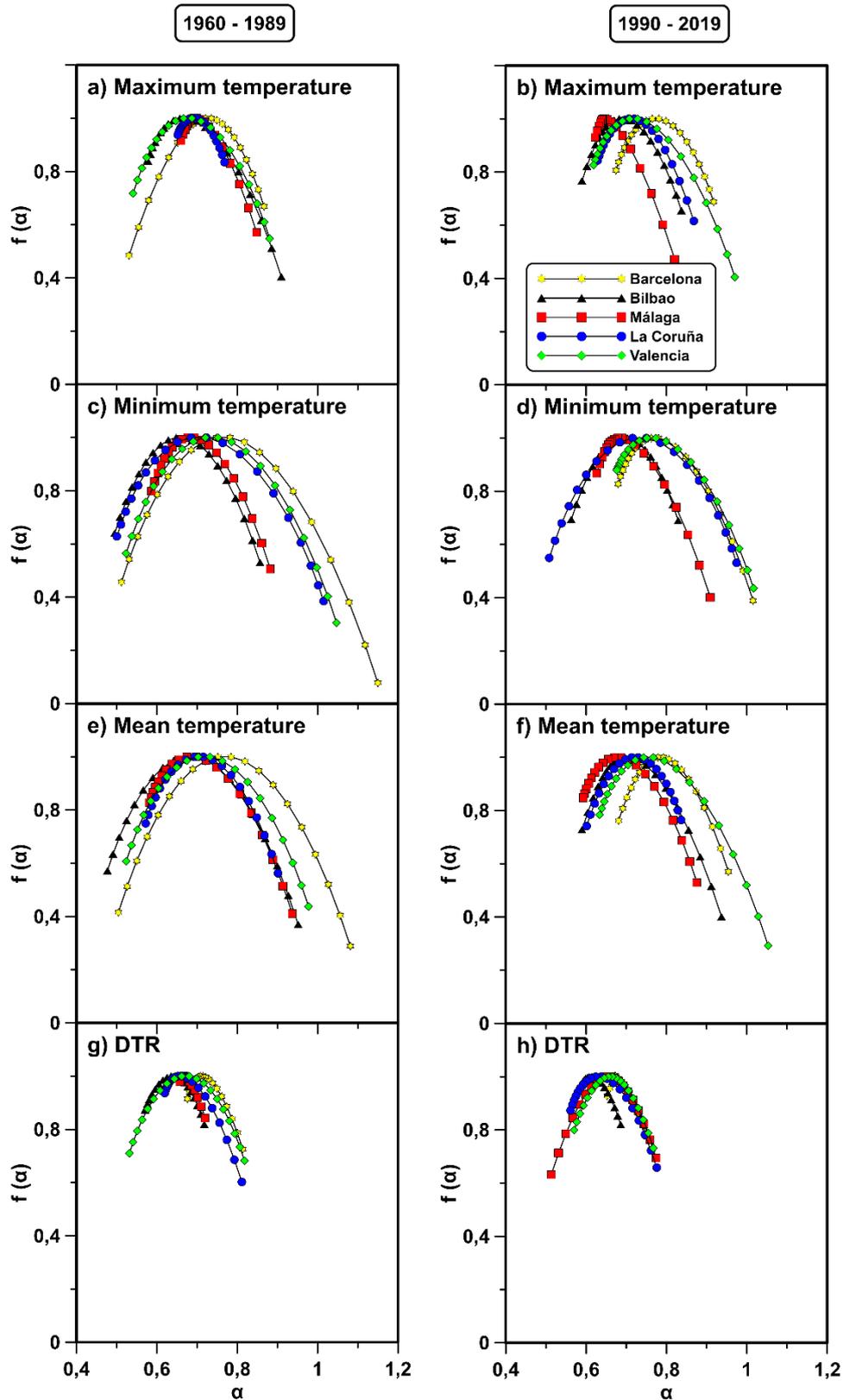

Fig. 7: Multifractal spectrum of $T_{max}$ (a and b), $T_{min}$ (c and d), $T_{mean}$ (e and f) and $DTR$ (g and h) for every coastal station (Barcelona, Bilbao, Málaga, La Coruña and Valencia). Charts on the left side are from the years 1960-1989 (a, c, e and g), while the ones on the right side correspond to the years 1990-2019.

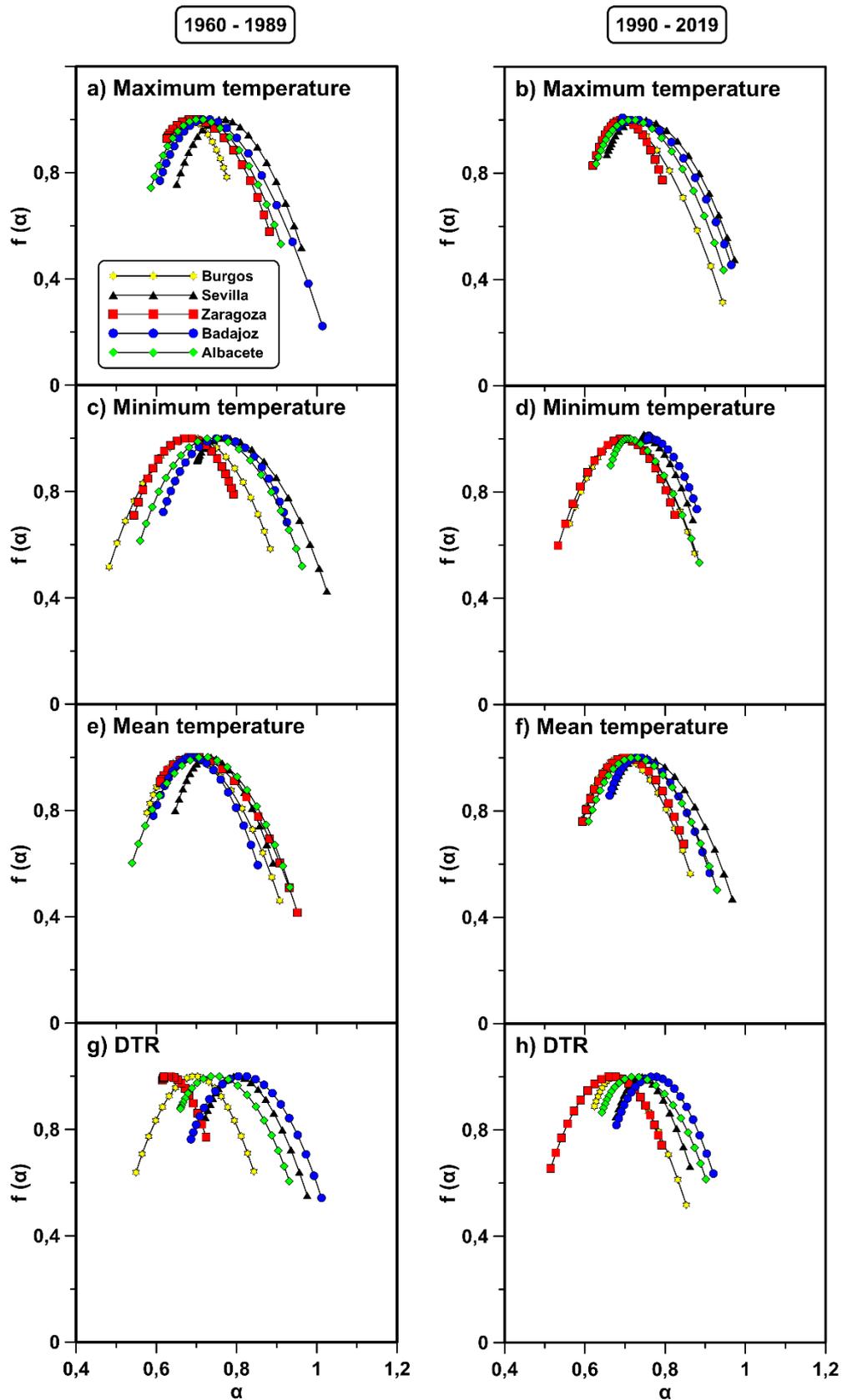

Fig. 8: Multifractal spectrum of $T_{max}$ (a and b), $T_{min}$ (c and d), $T_{mean}$ (e and f) and $DTR$ (g and h) for every mainland station (Burgos, Sevilla, Zaragoza, Badajoz and Albacete). Charts on the left side are from the years 1960-1989 (a, c, e and g), while the ones on the right side correspond to the years 1990-2019

| Station | $T_{max}$ | | $T_{min}$ | | $T_{mean}$ | | DTR | |
|---|---|---|---|---|---|---|---|---|
| | 1960 - 1989 | 1990 - 2019 | 1960 - 1989 | 1990 - 2019 | 1960 - 1989 | 1990 - 2019 | 1960 - 1989 | 1990 - 2019 |
| Barcelona | 0,336 | 0,243 | 0,638 | 0,336 | 0,577 | 0,273 | 0,137 | 0,099 |
| Bilbao | 0,332 | 0,247 | 0,362 | 0,266 | 0,475 | 0,348 | 0,147 | 0,098 |
| Málaga | 0,188 | 0,198 | 0,296 | 0,283 | 0,355 | 0,283 | 0,063 | 0,261 |
| La Coruña | 0,116 | 0,239 | 0,514 | 0,466 | 0,330 | 0,235 | 0,192 | 0,215 |
| Valencia | 0,340 | 0,351 | 0,524 | 0,340 | 0,454 | 0,419 | 0,286 | 0,197 |
| Burgos | 0,110 | 0,295 | 0,401 | 0,313 | 0,331 | 0,271 | 0,294 | 0,229 |
| Sevilla | 0,311 | 0,318 | 0,323 | 0,130 | 0,244 | 0,299 | 0,254 | 0,184 |
| Zaragoza | 0,255 | 0,172 | 0,248 | 0,291 | 0,343 | 0,253 | 0,109 | 0,277 |
| Badajoz | 0,405 | 0,282 | 0,308 | 0,126 | 0,260 | 0,249 | 0,325 | 0,241 |
| Albacete | 0,324 | 0,318 | 0,403 | 0,221 | 0,393 | 0,321 | 0,271 | 0,259 |

Table 1: Multifractal spectra width $W$ of daily maximum ($T_{max}$), minimum ($T_{min}$), mean temperature ($T_{mean}$) and diurnal temperature range ($DTR$) for the periods 1960-1989 and 1990-2019 in every station.

| Station | $T_{max}$ | | $T_{min}$ | | $T_{mean}$ | | DTR | |
|---|---|---|---|---|---|---|---|---|
| | 1960 - 1989 | 1990 - 2019 | 1960 - 1989 | 1990 - 2019 | 1960 - 1989 | 1990 - 2019 | 1960 - 1989 | 1990 - 2019 |
| Barcelona | -0,408 | 0,457 | 0,447 | 0,487 | 0,367 | 0,480 | 0,561 | 0,520 |
| Bilbao | 0,495 | 0,560 | 0,345 | -0,264 | 0,354 | 0,627 | -0,364 | 0,407 |
| Málaga | 0,294 | -0,218 | 0,516 | 0,425 | 0,472 | 0,493 | 0,004 | -0,336 |
| La Coruña | 0,432 | 0,413 | 0,487 | 0,361 | 0,418 | 0,284 | 0,345 | 0,498 |
| Valencia | 0,489 | 0,486 | 0,451 | 0,440 | 0,425 | 0,492 | -0,269 | -0,211 |
| Burgos | 0,221 | 0,085 | 0,248 | 0,411 | 0,541 | 0,519 | 0,233 | 0,421 |
| Sevilla | 0,508 | 0,455 | 0,377 | -0,147 | 0,620 | 0,452 | 0,516 | 0,518 |
| Zaragoza | 0,399 | 0,488 | 0,269 | -0,329 | 0,385 | 0,448 | 0,232 | -0,355 |
| Badajoz | 0,452 | 0,554 | 0,382 | 0,055 | 0,516 | 0,525 | 0,381 | 0,412 |
| Albacete | 0,519 | 0,488 | 0,432 | 0,414 | 0,399 | 0,521 | 0,447 | 0,492 |

Table 2: Asymmetry parameter $B$ of multifractal spectra of daily maximum ($T_{max}$), minimum ($T_{min}$), mean temperature ($T_{mean}$) and diurnal temperature range ($DTR$) for the periods 1960-1989 and 1990-2019 in every station.

## 4. CONCLUSIONS

The analyzed air surface temperature variables show all distinct scaling exponents when looking at the fluctuation functions ($F(q)$) versus scales ($s$) at different $q$ moments. This fact demonstrates the intrinsic multifractal nature of signals. It can be concluded that all the series are stationary and long-range correlated. A way to understand the long-range correlation is that an increase in temperature would be more likely followed by another increase and vice versa.

The main multifractal features of the four temperature signals vary between years 1960-1989 and 1990-2019, to a greater or lesser extent. This result might be interpreted as a possible relation between the climatic change and the fractal properties. However, in most cases, the symmetry of multifractal spectra remains almost the same between both periods and changes that we found lacked any consistency.

Regarding coastal locations, a higher degree of multifractality is mostly present in both periods in $T_{min}$ and $T_{mean}$, since they show wider spectra than for $T_{max}$ and $DTR$. This result is a discriminator element between coastal and mainland stations in both periods because the last ones do not show this pattern. Thus, authors conclude that ocean might have an impact on the higher complexity of minimum and mean temperature time series on these locations.

Nevertheless, a more relevant result obtained from $T_{min}$ and $T_{mean}$ is a spectral narrowing on the vast majority of mainland and coastal stations over time. This means that the complexity of the temperature series decreases. However, authors believe that changes in complexity for the mean might be derived from minimum temperature values. Since this effect is not consistent with the maximum temperature, there is an asymmetry in the temperature behavior. Brunet et al. already found an asymmetric behavior between maxima and minima only in mainland stations over the Iberian Peninsula in their statistical study between 1850 and 2003 [15]. In that study, maximum temperature increased at greater rates than minimum temperature. On the contrary, other investigations made by Esteban-Parra et al. in 2003 [16] or Staudt et al. in 2004 [17] obtained the opposite behavior (higher rates of change for minima than for maxima). According to this, the climatic change experienced in this region might be linked to different behaviors in maxima and minima. A relation between this asymmetry found in the Iberian Peninsula and the different multifractality shown by their singularity spectra could exist.

The conclusions drawn from these results can help testing models related to the climate change. One important point extracted from this analysis is that multifractal properties are not conserved over time for temperature time series. Hence, to improve future simulations, studies involving greater periods of time should be done. By doing so, a better understanding of how these parameters evolve with time could be achieved. Additionally, it must be pointed out the importance of seeking relations among the multifractal features and the atmospheric processes involved. A search of the applicability of these outcomes for assessing different climate models will be the aim of future works.

**ACKNOWLEDGEMENTS**

The FLAE approach for the sequence of authors is applied in this work. Authors gratefully acknowledge the support of the Andalusian Research Plan

Group TEP-957 and the Research Program of the University of Cordoba (2021), Spain. We also thank the Spanish Meteorological Agency ("Agencia Estatal de Meteorología") for providing data records.

# CREDIT AUTHOR STATEMENT

**Javier Gómez-Gómez:** Conceptualization, Methodology, Software, Validation, Formal Analysis, Data Curation, Investigation, Writing - Original Draft

**Rafael Carmona-Cabezas:** Conceptualization, Software, Investigation, Resources.

**Ana B. Ariza-Villaverde:** Conceptualization, Resources, Supervision.

**Eduardo Gutiérrez de Ravé:** Project Administration, Funding Acquisition, Supervision

**Francisco José Jiménez-Hornero:** Project Administration, Funding Acquisition, Supervision

Declaration of Interest Statement

**Declaration of interests**

☒ The authors declare that they have no known competing financial interests or personal relationships that could have appeared to influence the work reported in this paper.

☐ The authors declare the following financial interests/personal relationships which may be considered as potential competing interests:

# HIGHLIGHTS

- MF-DFA is used to study temperature across Spain in two subperiods of 30 years.
- Hurst exponents reveal that all series are long-range correlated and stationary.
- The width of spectra lessens in the last period for minimum and mean temperature.
- $T_{min}$ and $T_{mean}$ exhibit wider spectra than for $T_{max}$ and $DTR$ in coastal locations.

Graphical abstract

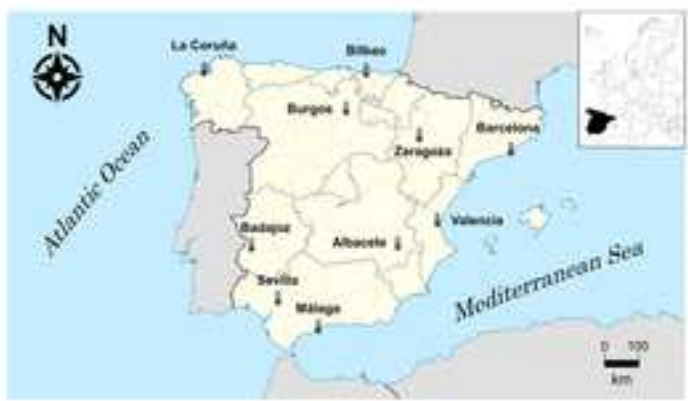
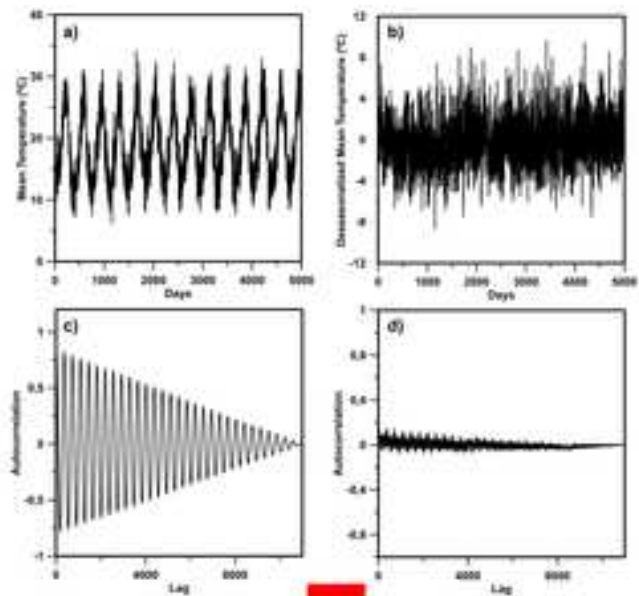
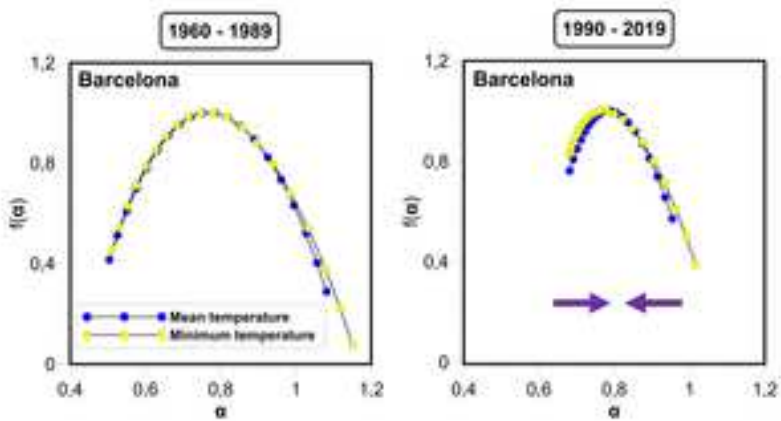